\documentclass[12pt]{article}
\usepackage{amsmath,amssymb,amscd}
\usepackage{epsfig}

\begin{document}

\begin{flushright}
{\bf FIAN/TD-09/05}\\{\bf ITEP-TH-38-05}
\end{flushright}

\bigskip
\bigskip

\begin{center}
{\bf ON COLLECTIVE NON-GAUSSIAN DEPENDENCE PATTERNS IN HIGH FREQUENCY FINANCIAL DATA}
\end{center}

\begin{center}
\bf{ \large Andrei Leonidov$^{(a,b,c)}$\footnote{Corresponding author. E-mail leonidov@lpi.ru}, Vladimir Trainin$^{(b)}$, \\Alexander
Zaitsev$^{(b)}$}
\end{center}
\medskip
(a) {\it Theoretical Physics Department, P.N.~Lebedev Physics Institute,\\
    Moscow, Russia}

(b) {\it Letra Group, LLC, 400 W. Cummings Park, Suite 3725, \\ Woburn MA 01801, USA}

(c) {\it Institute of Theoretical and Experimental Physics, Moscow, Russia}

\bigskip

\bigskip

\bigskip

\begin{center}
{\bf Abstract}
\end{center}
The analysis of observed conditional distributions of both lagged and simultaneous intraday price increments of a basket of stocks reveals phenomena
of dependence - induced volatility smile and kurtosis reduction. A model based on multivariate t-Student distribution shows that the observed effects
are caused by collective non-gaussian dependence properties of financial time series.

\newpage

\section{Introduction}

One of the fundamental problems of quantitative finance is to develop a description of collective dynamical properties of market prices of an
ensemble of financial instruments.

Let us stress that a problem of working out an economic description of the properties of market prices is not completely solved even at the level of
individual securities. A simple and very popular dynamical picture allowing transparent analytical treatment, that of a random walk, assumes a)
normal distribution for the price increments and b) independence of price increments corresponding to different time intervals. Starting from the
studies of Mandelbrot in the 60'th \cite{Man} through more recent analysis \cite{Lo,MS,BP,C01,LM00} there accumulated a large body of evidence that
real price dynamics for individual securities reveals substantial deviations from both assumptions.

Obviously we discover a much higher complexity when moving from a single financial security to a basket of securities. At this level we expect to
deal with such novel effects as a) specific non-gaussian properties of the multivariate distribution of price increments \cite{BP,LM00}; b) temporal
autocorrelations  in price changes of single securities \cite{BP,C01} mixed with simultaneous cross-correlations between price increments of
different basket ingredients \cite{BP,EG}. In fact, collective price dynamics is characterized by pronounced non-gaussian properties and a
complicated web of interdependencies.

In this study we apply a conditional distribution approach to scrutinize the dependence structure within an ensemble of financial instruments and the
related nongaussian effects. Analysis of the value of "response" conditioned on the "input" having a certain magnitude enables to explicitly quantify
the dependencies in the market data. Generically, dealing with a set of securities and following its temporal evolution, we can identify two types of
conditional distributions which are of interest to us: a) distribution of  a {\it future} price increment given that past price increments of all
securities lie in a certain range; b) distribution of a price increment given that all other price increments {\it in the same time interval} lie in
a certain range. A simple example of the phenomenon described by the former distribution is the lagged autocorrelation, of the latter - the
simultaneous cross-sectional correlation. Let us stress that separating time-lagged dependencies ("horizontal" for further reference) from
simultaneously existing ones ("vertical" for further reference) is a simplification of the generic picture which allows, however, to discuss various
types of dependencies in a simple setting. A generic dependence pattern is a "product" of both: past evolution of a subset of securities may
influence future evolution of another subset. An importance of these generic "non-diagonal" contributions was studied, in the context of
profitability of a simple contrarian strategy, in \cite{LM90}. Let us also mention the recent studies of lagged conditional distributions of daily
returns \cite{BM03,CJY05}, in the latter reference - in relation to a particular stochastic volatility model.

Analyzing, in terms of conditional distributions, the market data on intraday price increments of a large set of liquid stocks traded in NYSE and
NASDAQ we have found pronounced specific effects characterizing the conditional dynamics of price increments for both lagged and simultaneous types
of dependence. Most spectacular is a relationship between the volatility of the "response" increment and the magnitude of the "input" one which can
in simple terms be described as a dependence-induced volatility smile ("D"-smile).  Another striking feature seen in the data is a dramatic reduction
of the kurtosis of the conditional distribution of the "response" increments.

To give a quantitative interpretation of these results we have developed a model description of the corresponding conditional distributions based on
a multivariate non-gaussian t-Student distribution depending on both past and future price increments.  Let us note that a multivariate t-Student
distribution is a popular choice for analyzing the simultaneous \cite{BP} and lagged \cite{AFKS} correlations in financial dynamics. The non-gaussian
nature of the model turned out to be a key element enabling to explain the dependence structures observed in the market data. In particular,
conditional volatility smile and decrease of kurtosis take place even in complete absence of linear correlations. The above-described effects
completely disappear, however, if one uses a multivariate gaussian distribution depending on the corresponding matrix of covariances (correlations)
instead of the fat-tailed multivariate t-Student distribution.

\section{Observed features}

The object of our study is a dynamical evolution of a group of $N=100$ most liquid stocks from S\&P 500 \footnote{A list of stocks is given in the
Appendix} within a two-year time period from January 1, 2003 through December 31, 2004, characterized by the price increments $\delta p(\tau)$ in the
time interval of length $\tau$. In our analysis we use two intervals of length $\tau = 6 \, {\rm min}$ and $\tau = 60 \, {\rm min}$. For an interval
$[t,t+\tau]$ we thus have a configuration of $N$ price increments $\{ \delta p^j (t) \equiv p^j(t+\tau)-p^j(t) \}$, \,$j=1  \cdots N$, evolving in
time. Most interesting are, of course, the features of this evolution distinguishing it from that of a group of independent objects. Such cohesion
can be of both simultaneous (interrelations between the values of price increments of different stocks in the same time interval) and lagged
(interrelations between the price increments of the same or different stocks in different time intervals) nature.

Below we shall concentrate on the two simplest types of dependencies:
\begin{enumerate}
\item{Interrelation between the price increments in consecutive time intervals for the same stock ("horizontal" case)}
\item{Interrelations between the price increments of different stocks in the same time interval ("vertical" case)}
\end{enumerate}

Let us start with "horizontal" case and consider all pairs \{ $\delta p^j(t),\delta p^j(t+\tau) \}$ of stock price increments in two consecutive time
intervals for some given j-th stock. Our goal is to describe probabilistic properties of the set of increments at time $t+\tau$ conditioned on the
sign and magnitude of the increments at preceding time $t$. These properties are characterized by the corresponding conditional distribution
constructed as follows:
\begin{itemize}
 {\item
 First, we normalize the price increments $\delta p^j(t)$ in the first interval of the pair
 by their unconditional standard deviation $\sigma^j_{\rm tot}$, $\delta p^j(t) \rightarrow x^j(t) = \delta p^j(t)/ \sigma^j_{\rm tot}$
 }
 \item{
 Second, we divide the set of thus normalized increments into subintervals $\Delta_i$ having
 the fixed length $0.5$. The total interval we consider is $\Delta = [-3.25,3.25]$. The subinterval
 $\Delta_1$ thus corresponds (for j-th stock) to $x^j \in [-3.25,-2.75]$, etc.
 }
 \item{
 For a pair with $x^j$ belonging to some fixed subinterval $\Delta_i$ we study the conditional distribution ${\cal P}_{\Delta_i} (y^j)$
 of the normalized price increments $y^j = \delta p^j(t+\tau)/\sigma^j_{\rm tot}$ in the second interval of the pair
 \begin{equation}\label{condis0}
    {\cal P}_{\Delta_i} (y^j) \equiv {\cal P} (y^j |\, x^j \in \Delta_i)
 \end{equation}
 }
\end{itemize}
The distribution~(\ref{condis0}) is then a "horizontal" coarse-grained conditional distribution\footnote{Coarse graining refers to conditioned
variable $x$ belonging to some fixed interval $\Delta_i$: $x \in \Delta_i$}.

\begin{figure}[h]
\begin{center}
\epsfig{file=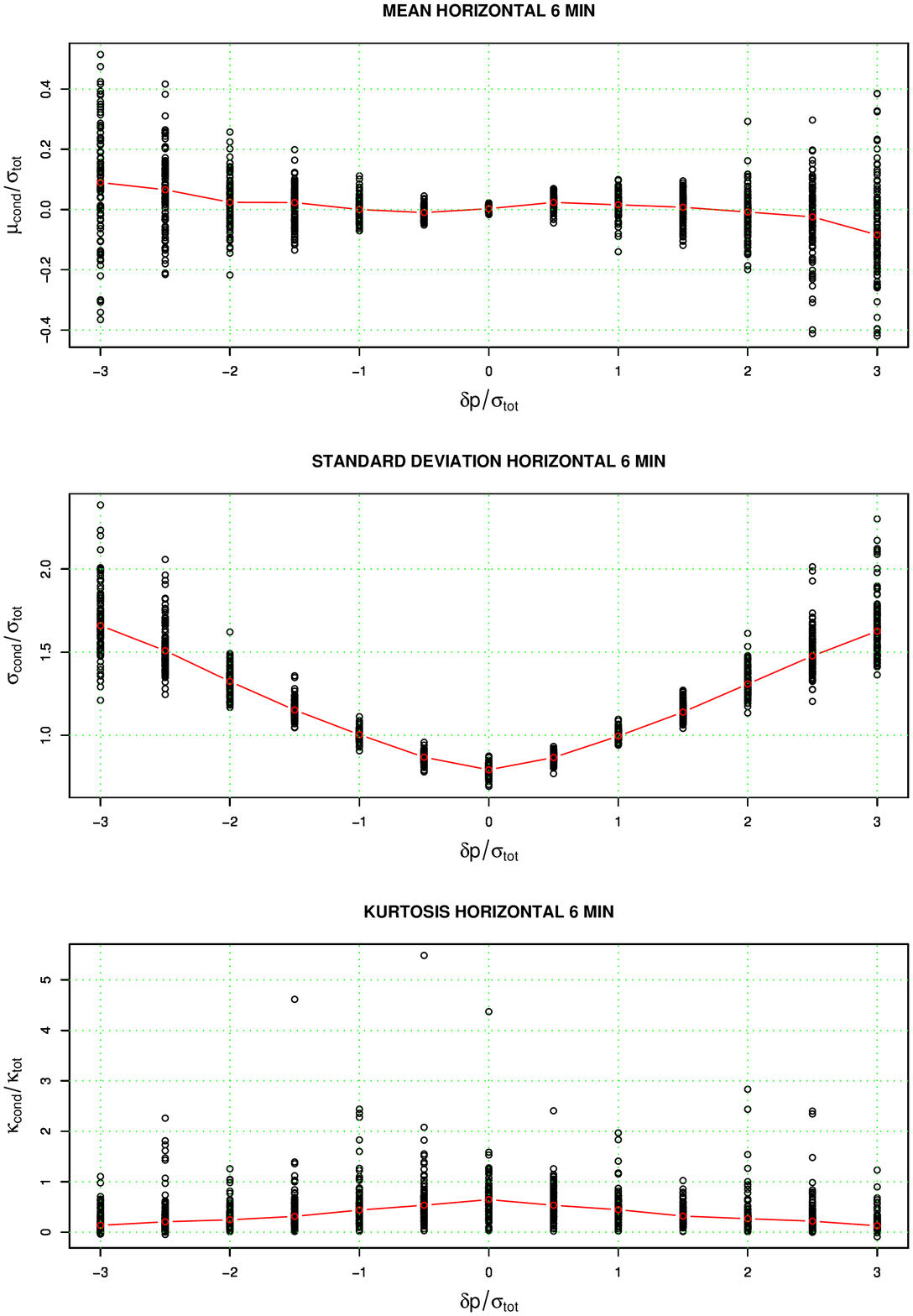,height=20cm,width=16cm}
\end{center}
\caption{Normalized mean, standard deviation and anomalous kurtosis of the coarse-grained "horizontal" conditional distribution versus the initial
push, $\tau=6 \, {\rm min}$, 100 stocks. Red lines show the medians of the scatterplots.} \label{shift1h}
\end{figure}

The basic properties of the conditional distribution ${\cal P}_{\Delta_i} (y)$  are conveniently summarized by the values of its lowest moments -
mean $\mu_{\rm cond}$, standard deviation $\sigma_{\rm cond}$, anomalous kurtosis $\kappa_{\rm cond}$, etc. . In this paper we shall study the
correspondingly normalized conditional mean, conditional standard deviation and conditional anomalous kurtosis. The above-described normalization
allows to consider all stocks simultaneously. The normalized mean $\mu_{{\rm cond}}/\sigma_{{\rm tot}}$, standard deviation $\sigma_{{\rm
cond}}/\sigma_{\rm tot}$ and anomalous kurtosis $\kappa_{{\rm cond}}/\kappa_{{\rm tot}}$, where $\kappa_{{\rm tot}}$ is an unconditional anomalous
kurtosis of the increments' distribution, of the "horizontal" coarse-grained conditional distribution (\ref{condis0}) (i.e. that characterizing the
set of all adjacent 6-min. intervals for each stock) are plotted as a function of the rescaled initial push $\delta p / \sigma_{{\rm tot}}$ in
Fig.~\ref{shift1h}\,.

Let us now turn to the analysis of the "vertical" interrelations between simultaneous price increments of different stocks The corresponding
coarse-grained conditional distribution is constructed in complete analogy with the above-described "horizontal" case:
 \begin{equation}\label{condis01}
    {\cal P}_{\Delta_i} (y^j) \equiv {\cal P} (y^j |\, x^k \in \Delta_i)\,,
 \end{equation}
where the conditioned variable $x^k \equiv \delta p^k (t)/\sigma^k_{\rm tot}$ refers to the $k$-th stock, and the response variable $y^j \equiv
\delta p^j (t)/\sigma^j_{\rm tot}$ - to the $j$-th one.

\begin{figure}[h]
\begin{center}
\epsfig{file=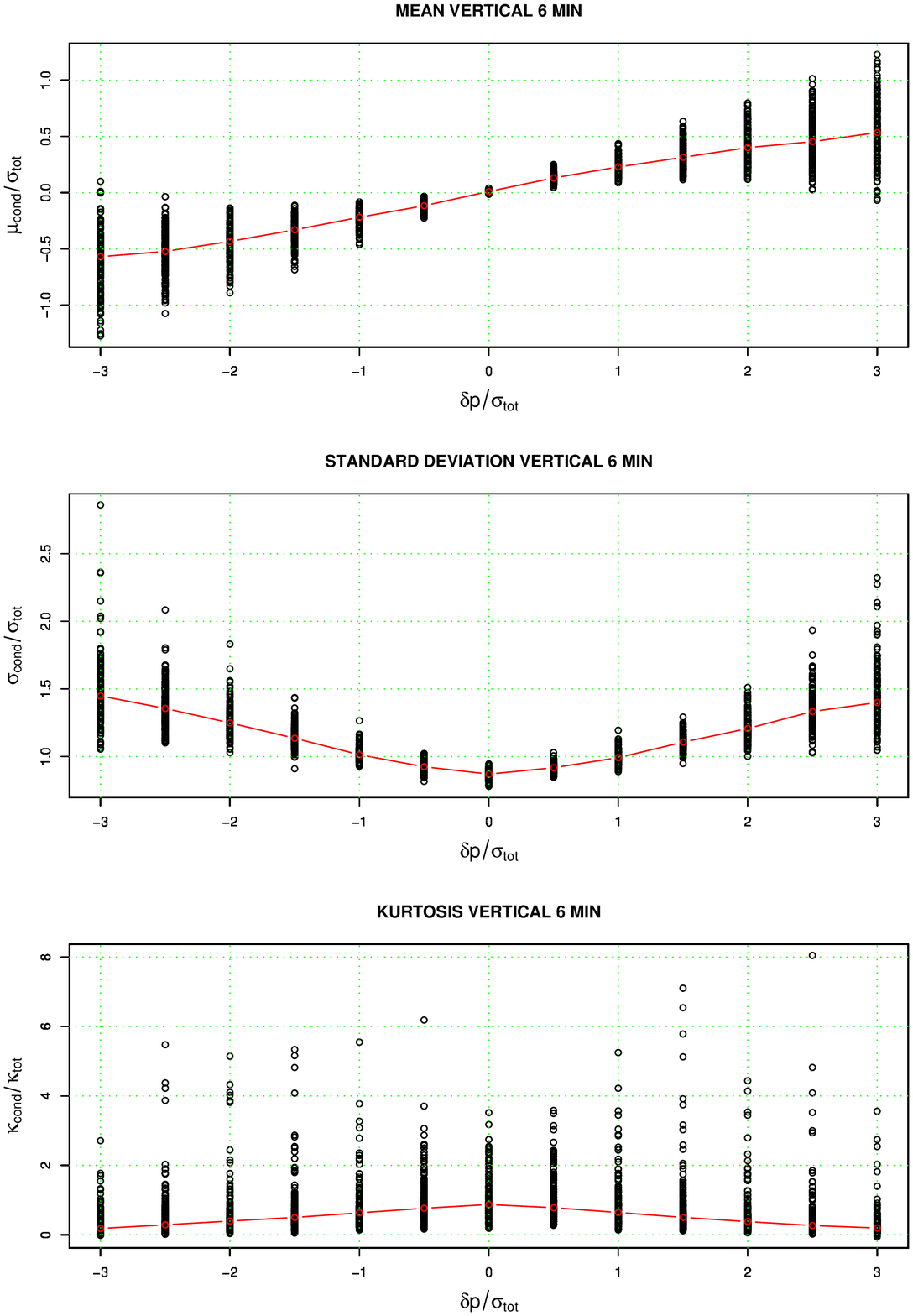,height=20cm,width=16cm}
\end{center}
\caption{Normalized mean, standard deviation and anomalous kurtosis of the coarse-grained "vertical" conditional distribution  versus the initial
push, $\tau=6 \, {\rm min}$ , 100 stocks. Red lines show the medians of the scatterplots.} \label{shift1v}
\end{figure}

 In Fig.~\ref{shift1v} we show the normalized conditional mean, standard deviation and kurtosis for
6-min. intervals for the "vertical"case.

\begin{figure}[h]
\begin{center}
\epsfig{file=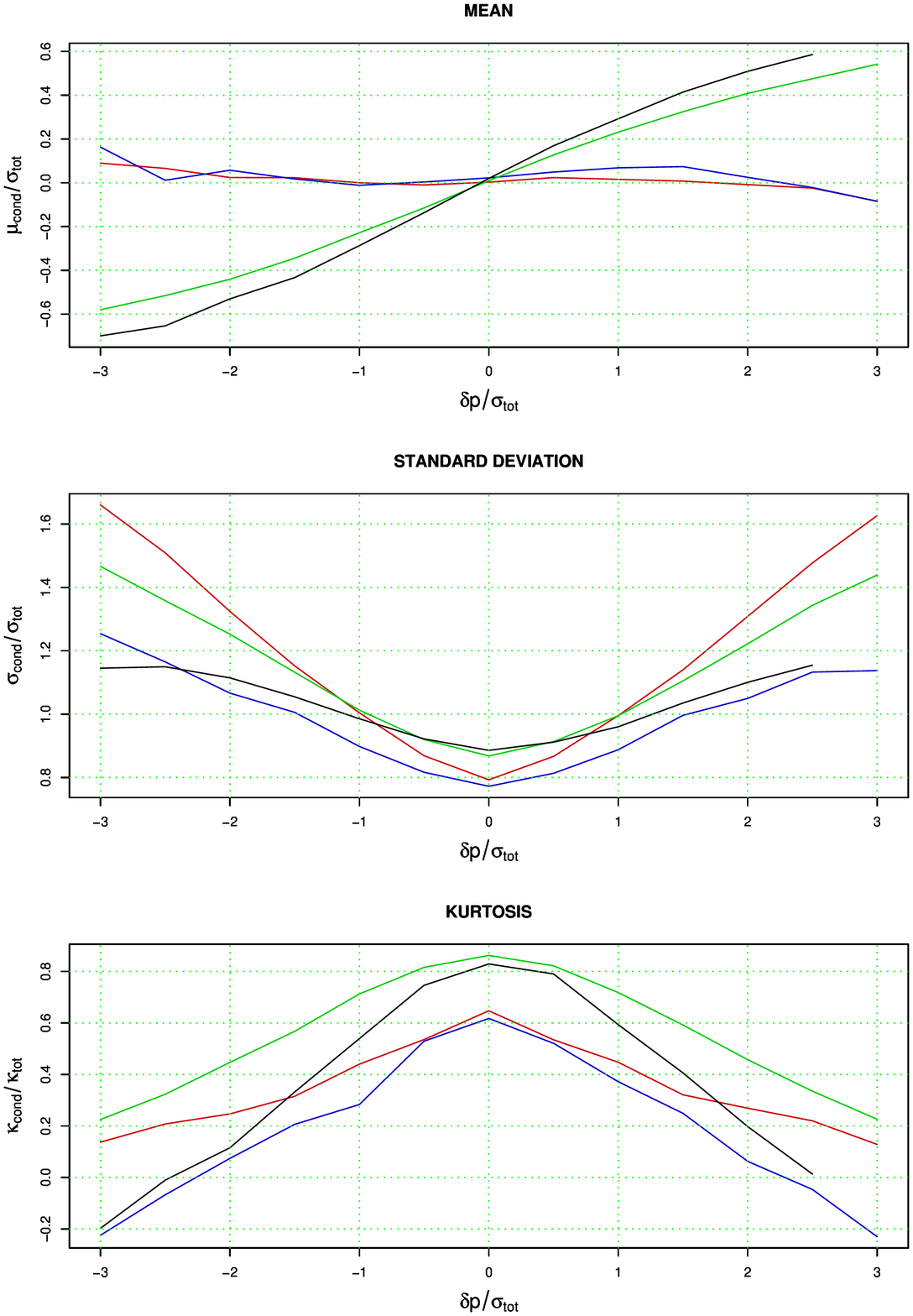,height=20cm,width=16cm}
\end{center}
\caption{Medians of normalized mean, standard deviation and anomalous kurtosis of the coarse-grained conditional distribution versus the initial
push, $\tau=6 \, {\rm min}$ ("horizontal": red line, "vertical": blue line) and $\tau=60 \, {\rm min}$
 ("horizontal": green line, "vertical": black line). }
\label{allmed}
\end{figure}

In Fig.~\ref{allmed} we plot the medians of the scatterplots for the normalized conditional mean, standard deviation and kurtosis for 6-min. and
60-min. intervals, combining the "horizontal" and "vertical" quantities.

The analysis of Figs.~\ref{shift1h}, \ref{shift1v} and \ref{allmed} leads to the following conclusions:
\begin{itemize}
 \item{
      The resulting plots for conditional mean $\mu_{{\rm cond}}$ in the "horizontal" case are too noisy to allow unambiguous
      interpretation. In the "vertical" case one observes, for both cases of $\tau = 6 \, {\rm min}$ and
      $\tau = 60 \, {\rm min}$, a picture consistent with that of conditional
      mean generated through the presence of positive correlation, see below Eqs.~(\ref{condmoms}) and (\ref{conpar2}).
      }
 \item{
      The plots of the relative conditional standard deviation $\sigma_{{\rm cond}}$ in "horizontal" and "vertical" case
      are, for the both cases of $\tau = 6 \, {\rm min}$ and $\tau = 60 \, {\rm min}$, strikingly similar.
      For $\tau = 6 \, {\rm min}$ we observe a pronounced conditional volatility smile, or dependence-induced volatility smile (D-smile)
      (see a more detailed discussion of this phenomenon in the next section),
      such that at small $x$ the standard deviation of the response is smaller than the unconditional
      standard deviation, while in the tails it is, on contrary, larger. For $\tau = 60 \, {\rm min}$
      the smile is noticeably flatter than for  $\tau = 6 \, {\rm min}$. This effect can be explained by the
      decay of anomalous kurtosis of price increments with growing $\tau$, see below discussion after Eq.~(\ref{conpar2}).
      }
 \item{
      The median conditional kurtosis is noticeably smaller than the unconditional one.
      }
\end{itemize}

We see, that in both "vertical" and "horizontal" cases the data shows, for both scales of $\tau = 6 \, {\rm min}$ and $\tau = 60 \, {\rm min}$, the
same rather nontrivial patterns: conditional volatility smile and decrease of conditional kurtosis. The origin of the first effect is discussed in
the next section. We shall argue, that it is in the probabilistic dependence of the adjacent price increments, whereas the role of linear correlation
effects is in fact minor.

\section{Model}

Let us now present a model that explains the phenomenona of dependence-induced volatility smile and kurtosis reduction in the coarse-grained
conditional distributions described in the previous section.

At the fundamental level of description the model describing the behavior of $N$ securities in two adjacent time intervals is fully specified by a $2
N$ - dimensional probability distribution. The focus of our study is on the properties of the conditional distributions constructed from this basic
enveloping $2N$-dimensional distribution. Generically conditional distributions are obtained by restricting the values of a subset of variables. Let
us collectively denote these variables by ${\bf x}$, where ${\bf x}$ is a $N_{\bf x}$ - dimensional vector. Generically the vector ${\bf x}$ can
include increments belonging to different time intervals. We are thus dealing with a conditional distribution depending on $N_{\bf y} \equiv 2N -
N_{\bf x}$ variables. If we stay within the class of elliptical distributions, the multivariate probability distribution is a function of a quadratic
form ${\cal K}$ constructed from the vector ${\bf z}^\top = ({\bf y},{\bf x})$ and the generalized covariance matrix $\Sigma$, ${\cal K} = {\bf
z}^\top \cdot \Sigma^{-1} \cdot {\bf z}$. The covariance matrix $\Sigma$ includes the $N_{\bf x} \times N_{\bf x}$ covariance matrix $C_{\bf x}$
describing the correlations within the subset of conditioned variables ${\bf x}$, the $N_{\bf y} \times N_{\bf y}$ covariance matrix $C_{\bf y}$
describing the correlations within the subset of the variables ${\bf y}$  and the $N_{\bf x} \times N_{\bf y}$ covariance matrix $C_{\bf xy}$
describing the cross-covariances between the two groups:
\begin{equation}\label{Sigma}
 \Sigma =
\left(
\begin{array}{cc}
  C_{\bf y} & C_{\bf xy} \\
  C_{\bf xy}^\top & C_{\bf x}
\end{array}
\right)
\end{equation}

At this stage we have to give an explicit description of the multivariate distribution containing the covariance matrix $\Sigma$. As will be
elucidated below, a simplest choice of a gaussian multivariate distribution does not allow to explain the phenomena of D-smile and kurtosis
reduction. There is, therefore, a clear need of taking into account the non-gaussian effects. The simplest possibility of keeping a fat-tailed nature
of the probability distributions of individual increments is to construct a multivariate distribution from the fat-tailed marginals. Recombination of
these marginals into a multivariate distribution requires constructing an appropriate copula. This construction is not unique, so the choice is
guided by simplicity and ability to reproduce basic features of market data \cite{MS03,BDE03}. In what follows we will show that a multivariate
t-Student distribution makes a good job in this respect, while the Gaussian multivariate distribution  fails to reproduce the properties of
conditional distributions observed in market data.

Let us consider a $2N$-dimensional t-Student distribution
\begin{equation}\label{tsdis}
 P^{(2N)}_S \, = \, \frac{1}{\sqrt{(\pi \mu)^{2N} \xi_{\mu}^{2N} {\rm det} \Sigma}}  \,
  \frac{\Gamma \left( \frac{\mu+2N}{2} \right)}{\Gamma \left( \frac{\mu}{2} \right)}
  \left[ 1+\frac{1}{\mu} \frac{1}{\xi_{\mu}}
 {\bf z}^\top \, \Sigma^{-1}\, {\bf z}\right]^{-\frac{\mu+2N}{2}}.
\end{equation}
where $\xi_{\mu} = (\mu-2)/\mu$ is a normalization factor ensuring, in particular, that the covariances computed with the distribution (\ref{tsdis})
are equal to the corresponding matrix elements of the matrix $\Sigma$.

Fixing some particular configuration of the "initial" increments ${\bf x}={\bf x_0} $ leads to the conditional distribution
(see, e.g., \cite{BJ70}):
\begin{eqnarray}\label{tsdiscon}
P^{(N)}_S({\bf y |\, x_0}) & = & \frac{1}{\sqrt{(\pi (\mu+N_{\bf x}))^{N_{\bf y}} \xi_{\mu+N_{\bf x}}^{N_{\bf y}} {\rm det \Sigma_{{\bf y|\,x_0}}}}}
\,
 \frac{\Gamma \left( \frac{\mu+N_{\bf x}+N_{\bf y}}{2} \right)}{\Gamma \left( \frac{\mu}{2} \right)} \nonumber \\
 & \times & \left[ 1+\frac{1}{\mu+N_{\bf x}} \frac{1}{\xi_{\mu+N_{\bf x}}}
 ({\bf y}-\langle {\bf y} \rangle_{\bf x_0})^\top \, \Sigma_{\bf y|\,x_0}^{-1} \,
  ({\bf y}-\langle {\bf y} \rangle_{\bf x_0}) \right]^{-\frac{\mu+N_{\bf x}+N_{\bf y}}{2}}.
\end{eqnarray}
The conditional distribution (\ref{tsdiscon}) is a multivariate $N_{\bf y}$ - dimensional t-Student distribution with the index $\mu+N_{\bf x}$ and
the following expected mean and covariance matrix:
\begin{eqnarray}\label{condmoms}
\langle {\bf y} \rangle_{\bf x_0} & = & C_{\bf xy} C_{\bf x}^{-1} \, {\bf x_0} \nonumber \\
  \Sigma_{\bf y|\,x_0} & = &
  \left( C_{\bf y} - C_{\bf xy} C_{\bf x}^{-1} C_{\bf xy}^\top \right)
  \left[ \frac{\mu-2}{\mu+N_{\bf x}-2} \right]\,
  \left[ 1+\frac{1}{\mu} \frac{1}{\xi_{\mu}} {\cal K}_{\bf x_0} \right]\, ,
\end{eqnarray}
where $ {\cal K}_{\bf x_0} = {\bf x_0}^\top C_{\bf x}^{-1} {\bf x_0}$. Let us note, that if we had used the Gaussian multivariate distribution for
constructing the conditional distribution analogous to (\ref{tsdiscon}), we would obtain a Gaussian conditional distribution with the following
expected mean and covariance matrix:
\begin{eqnarray}\label{condmomg}
\langle {\bf y} \rangle_{\bf x_0}^G & = & C_{\bf xy} C_{\bf x}^{-1} \, {\bf x_0} \nonumber \\
  \Sigma_{\bf y|\,x_0}^{G} & = & C_{\bf y} - C_{\bf xy} C_{\bf x}^{-1} C_{\bf xy}^\top
\end{eqnarray}
Comparing Eqs.~(\ref{condmoms}) and (\ref{condmomg}) we see, that the expected mean is in both cases the same, whereas the expected variance in the
t-Student case is a product of the gaussian expression and a $\mu$- and ${\cal K}_{\bf x}$ - dependent factor. An additional important phenomenon in
the case of a t-Student distribution is an increase of the tail exponent determining the fat-tailedness of the distribution: $\mu \Rightarrow
\mu+N_{\bf x}$ that thereby reduces the anomalous kurtosis\footnote{Note that the extent of this "gaussization" depends on the number of conditioned
variables which in the considered example is equal to $N_{\bf x}$. }
\begin{equation}
\kappa = \frac{6}{\mu-4} \,\,\,\,\, \Longrightarrow \,\,\,\,\, \kappa = \frac{6}{\mu+N_{\bf x}-4}
\end{equation}

To describe the conditional volatility smile phenomenon discussed in the previous section, one clearly needs initial conditions' depending
covariances. From the formula (\ref{condmomg}) we see that in the Gaussian case this effect is absent, whereas for t-Student distribution the
required dependence is manifest (see the second expression in (\ref{condmoms}) containing the factor of $ \left[ 1+\frac{1}{\mu} \frac{1}{\xi_{\mu}}
{\cal K}_{\bf x_0} \right]$). Of course, one should still prove that this dependence allows to describe the market data, see below. Nevertheless,
already at this stage of our analysis, one can conclude that the phenomenon of conditional volatility smile can be explained only by non-gaussian
effects - simply because the gaussian formalism does not have room for its description.

The conditional distribution Eq.~(\ref{tsdiscon}) summarizes the impact the "initial" configuration ${\bf x_0}$ has on the "final" one ${\bf y}$.

\begin{figure}[h]
\begin{center}
\epsfig{file=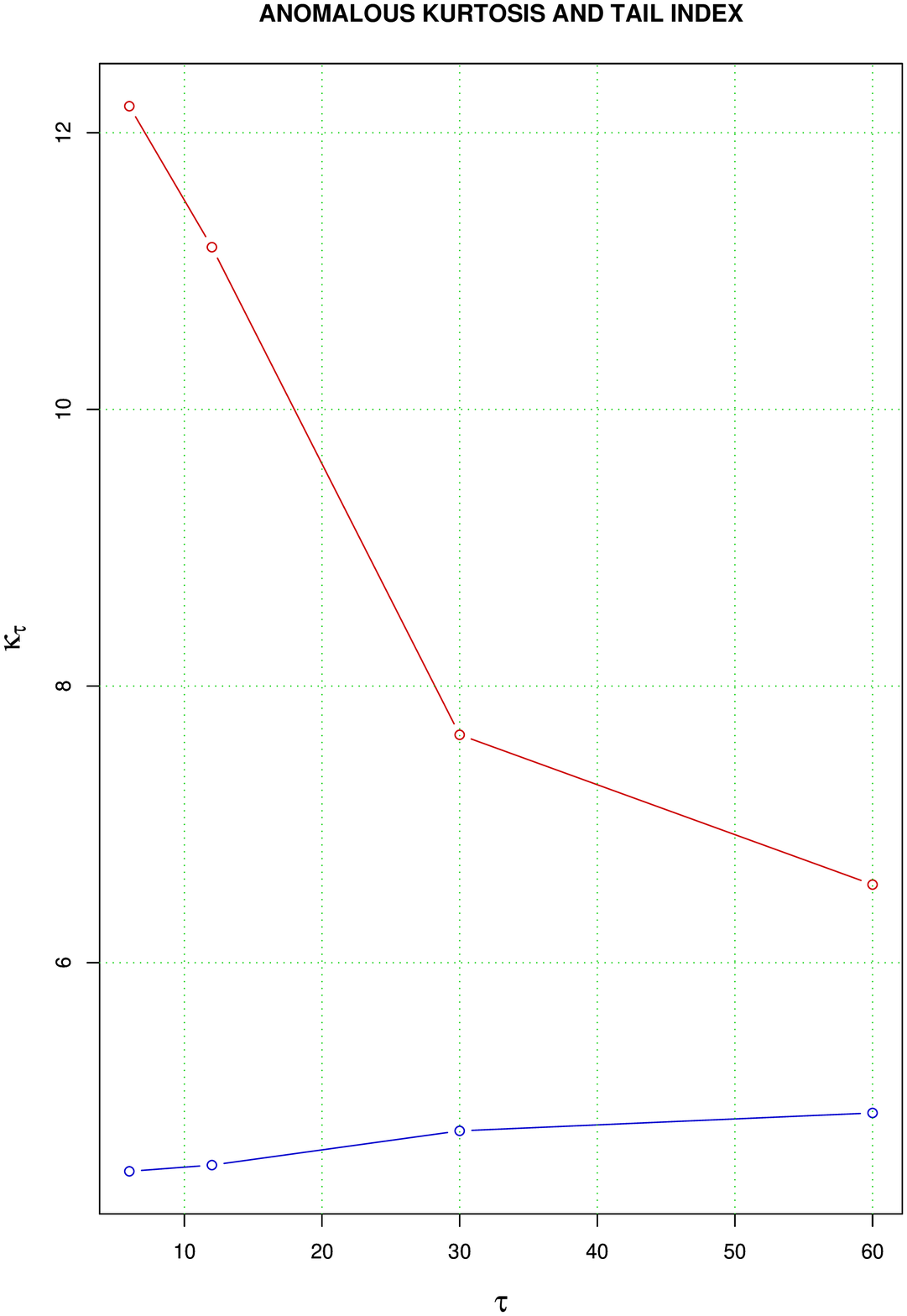,height=20cm,width=16cm}
\end{center}
\caption{Unconditional anomalous kurtosis $\kappa_\tau$ (red line) and corresponding tail index of the t-Student distribution $\mu_\tau$ (blue line)
for several intraday time intervals $\tau$.} \label{kamu}
\end{figure}

The explanation of the conditional volatility smile and kurtosis reduction effects described  in the previous section requires a simpler
$2$-dimensional version of (\ref{tsdis}) with one-dimensional $y$ and $x$. Let us thus consider two price increments in the two consecutive time
intervals for the same stock for the "horizontal" case (or the simultaneous increments of two stocks for the "vertical" case) and introduce the
corresponding bivariate distribution
\begin{equation}\label{tdis2}
 P^{(2)}_S(y,x) \, = \, \frac{1}{\sqrt{(\pi \mu)^2 \xi_{\mu}^2 {\rm det} \Sigma}}
\frac{\Gamma \left( \frac{\mu+2}{2} \right)}{\Gamma \left( \frac{\mu}{2}\right)}
\frac{1}{\left( 1 + \frac{1}{\mu} \frac{1}{\xi_\mu} K_\Sigma (x,y) \right)^{\frac{\mu+2}{2}}}
\end{equation}
Here $\Sigma$ is a covariance matrix
\begin{equation}
 \Sigma \, = \,
\begin{pmatrix}
   \sigma_y^2 & \sigma_x \sigma_y r \\
   \sigma_x \sigma_y r & \sigma_x^2
\end{pmatrix}
\end{equation}
and $K_\Sigma = (y,x) \cdot \Sigma^{-1} \cdot (y,x)^\top$ The conditional distribution ${\cal P} (y | \, x=x_0)$ corresponding to the above
distribution is again a t-Student distribution with the tail exponent $\mu+1$, conditional mean $\langle y \rangle_{x_0}$ and conditional $x_0$ -
dependent variance $\sigma^2_{y|x_0}$
\begin{eqnarray}\label{conpar2}
 \langle y \rangle_{x_0} & = & r\,x_0 \nonumber \\
 \sigma_{y|x_0}^2 & = & \sigma_y^2 (1-r^2)\, \frac{\mu-2}{\mu-1}
 \left(1+\frac{1}{\mu} \frac{1}{\xi_\mu} \frac{x_0^2}{\sigma_x^2} \right)
\end{eqnarray}
Therefore the conditional distribution is more gaussian (the ratio of its anomalous kurtosis to the unconditional one is equal to $(\mu-4)/(\mu-3) <
1$), but its standard deviation can be smaller or larger than the unconditional value $\sigma_y$ depending on the value of the conditioned variable
$x_0$. The parabolic dependence of the conditional volatility on the initial push $x_0$ is just the feature we need to explain the D-smiles in
Figs.~\ref{shift1h}, \ref{shift1v} and \ref{allmed}. The fine structure we have observed -- namely, the flattening of the D-smile with growing
$\tau$, can also be explained with the help of Eq.~(\ref{conpar2}). Indeed, the coefficient at $x_0^2$ is equal to $1/(\mu \xi_\mu) \equiv
1/(\mu-2)$. Now the data shows (see below Fig.~\ref{kamu}) that for larger time intervals the unconditional anomalous kurtosis $\kappa_\tau$ is
smaller, and the tail index $\mu_\tau = 4 + 6/\kappa_\tau$ is, correspondingly, larger, leading to the desired flattening of the smile. The
unconditional anomalous kurtosis $\kappa_\tau$ and the corresponding tail index $\mu_\tau$ are plotted for the ensemble of $N=100$ stocks considered
in the paper for several intraday time intervals, in Fig.~\ref{kamu}.

Let us also note that in the gaussian case one has $\langle y \rangle_x = r \cdot x_0$ and $\sigma_{y|x_0}^2 = \sigma_y^2 (1-r^2)$ so, as has been
already mentioned, the gaussian probabilistic link between the price increments does not leave room for $x_0$ - dependent effects in the conditional
covariance matrix.

\begin{figure}[h]
\begin{center}
\epsfig{file=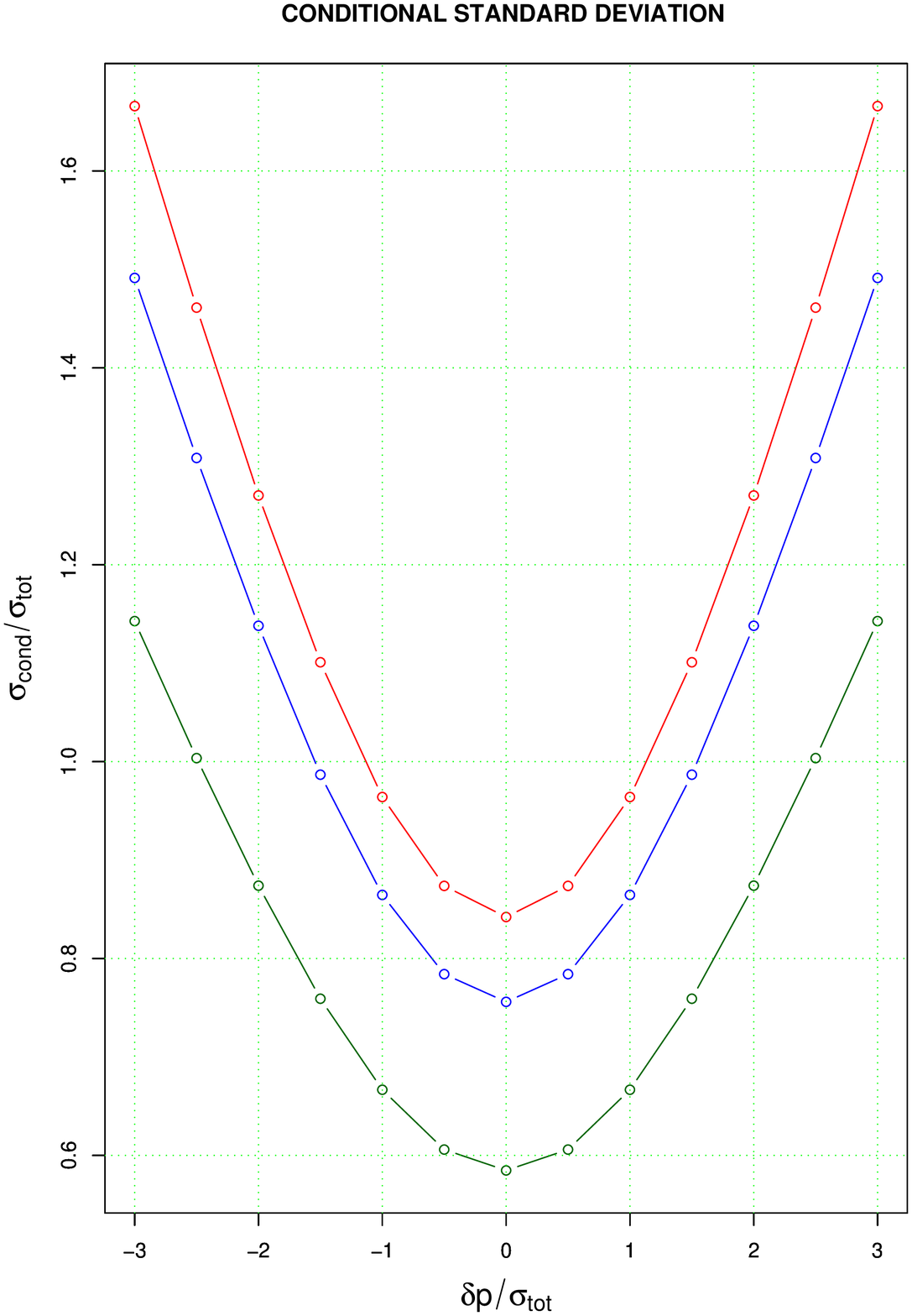,height=20cm,width=16cm}
\end{center}
\caption{Normalized standard deviation of the model coarse-grained conditional distribution versus the initial push; red: $r=0.25$, blue: $r=0.5$,
 green; $r=0.75$}
\label{model}
\end{figure}

To make the correspondence with the market data quantitative we should, however, introduce a coarse-grained version of the
conditional distribution ${\cal P} (y |\, x \in \Delta )$, where the variable $x$ belongs to a certain subinterval
$\Delta$:
\begin{equation}
 {\cal P} (y |\, x \in \Delta ) \, = \,
 \frac{\int_{x \in \Delta} dx \, P^{(2)}_S (y,x)}{\int_{x \in \Delta} dx \, P_S^{(1)} (x)}
\end{equation}
We have computed the normalized mean, relative standard deviation and anomalous kurtosis of a set of conditional distributions corresponding to the
same coarse-graining of the increments $x \equiv \delta p(t)/\sigma_{\rm tot}$ as used in the analysis of the market data in the previous section,
tail index $\mu=5$ and a set of correlation coefficients $r=0.25,0.5,0.75$. The conditional mean is, of course, simply proportional to $x_0$. The
conditional kurtosis drops to the expected $\kappa=3$, with small deviations. Most interesting is, of course, the behavior of the conditional
standard deviation shown in Fig.~\ref{model}.
We see that the model reproduces the conditional volatility smile with characteristics very similar to those observed
in the market data.

A crucial point in the correct interpretation of the above result is that linear correlation (present through the correlation coefficient $r$) shows
itself only via setting the absolute scale for the variance, see the second of Eq.~(\ref{conpar2}). It is clear,that the conditional volatility smile
would be present even in the complete absence of correlations ($r=0$). Therefore it is really appropriate to call the volatility dependence in
question a dependence-induced volatility smile (D-smile). Considering for instance the "horizontal" case, the probabilistic dependence between the
increments $\delta p(t)$ and $\delta p(t+1)$ can be manifestly demonstrated by computing, e.g., the correlator of their absolute values $G(1) =
\langle |\, \delta p (t)| \cdot |\, \delta p (t+1)| \rangle_t -( \langle |\, \delta p | \rangle_t )^2)$. Calculating this correlator for the
bivariate t-Student distribution (\ref{tdis2}) and its Gaussian counterpart gives
\begin{eqnarray}
 G_G(1) & = & \frac{2}{\pi} \, \sigma_{\rm tot}^2 \,r^2 \, \left(\sqrt{1-r^2}+\frac{\rm{Arcsin} \, r}{r} \right) \label{GG} \\
 G_S(1) & = & \frac{\mu}{\pi}\,\sigma_{\rm tot}^2
 \left[
 \frac{
       \Gamma \left( \frac{\mu-2}{2} \right) \Gamma \left( \frac{\mu}{2} \right) -
        \Gamma^2 \left( \frac{\mu-1}{2} \right)
      }{\Gamma^2 \left( \frac{\mu}{2} \right)}
 \right] + \nonumber \\
 & & \frac{\mu}{\pi}\,\sigma_{\rm tot}^2
     \frac{\Gamma \left( \frac{\mu-2}{2} \right)}{\Gamma \left( \frac{\mu}{2} \right)} \,
     r^2 \, \left[\sqrt{1-r^2}+\frac{\rm{Arcsin} \, r}{r} \right] \label{GS}
\end{eqnarray}
In the gaussian case the (linearly) uncorrelated variables are also independent and, indeed, the correlator (\ref{GG}) vanishes as $r^2$ at $ r \to
0$. In the case of t-Student distribution the correlator (\ref{GS}) is, on contrary, nonzero at $r = 0$, so increments are in this case
probabilistically dependent. Let us stress that this dependence is in fact imposed by the form of the unconditional distribution we have chosen. One
crucial feature is that the t-Student distribution ensures, in agreement with observations, that the corresponding marginal distributions are
fat-tailed. The t-Student copula we have used provides a framework in which the dependence effects are present even in the complete absence of linear
correlations.

\section{Discussion}

There still remains a number of important issues related to the questions discussed in the paper that we leave for the future analysis \cite{LTZ}.

First, one would like to generalize the binary-level description of simultaneous "vertical" interdependence of stock price increments to the fully
multivariate case of the influence of the $n$-point "trigger" configuration $\{ \delta p_j (t) \}$ on the move of the $k$-th stock in the next time
interval $\delta p_j (t+\tau)$.

Second, perhaps more difficult issue is studying the properties of the conditional distributions for arbitrary separation of corresponding time
intervals. Preliminary analysis of the market data shows the dependence of D-smile on this separation ("maturity"). This is to be expected from the
fact that volatility autocorrelations decay, albeit slowly, with time. This forces to generalize the formalism we have used\footnote{For an example
of a construction of this sort see \cite{AFKS}.}. In any case, a big goal is to establish connection with the explicit models of volatility dynamics,
see e.g. \cite{BL01,GZ03}, including the leverage effects \cite{BP}.

Finally, we would like to analyze in more details application of the nonlinear patterns we have described to portfolio optimization problems.
Expected mean, volatility and degree of fat-tailedness are crucial ingredients of portfolio optimization schemes \cite{BP,EG}, so specific effects
related to them are of clear interest in this context.

\section{Conclusion}

Let us summarize the main results of the present paper.

The focus of our analysis is on the properties of conditional distributions characterizing the probabilistic behavior of an ensemble of financial
instruments. The analysis of market data in the simplest case of a binary probabilistic dependence has revealed two major effects:
\begin{itemize}
 \item{The smile-shaped dependence of conditional volatility on the magnitude of the input due to
        non-gaussian nature of the enveloping t-Student distribution}
 \item{A noticeable reduction of the conditional anomalous kurtosis as compared to the unconditional one}
\end{itemize}
Let us also mention the flattening of the D-smile with growing time interval on which the price increments are computed.

We have constructed an explicit model characterizing the collective probabilistic pattern of an ensemble of price increments that gives a natural
explanation of the above-listed phenomena. The model is based on a multinomial t-Student distribution. This theoretical framework allows to
unambiguously relate the effects of a dependence-induced volatility smile and kurtosis reduction to the non-gaussian nature of the eneveloping
distribution.

\begin{center}
{\bf Acknowledgements}
\end{center}

The authors are very grateful to Eugene Pinsky for discussions and comments.

The work of A.L. was supported by the RFBR Grant 04-02-16880, and the Scientific school support grant 1936.2003.02

\section{Appendix}

Below we give a list of stocks studied in the paper:

\medskip

 A, AA, ABS, ABT, ADI, ADM, AIG, ALTR, AMGN, AMD, AOC, APA, APOL, AV, AVP, AXP,
 BA, BBBY, BBY, BHI, BIIB, BJS, BK, BLS, BR, BSX,
 CA, CAH, CAT, CC, CCL, CCU, CIT, CL, COP, CTXS, CVS, CZN,
 DG, DE,
 EDS, EK, EOP, EXC,
 FCX, FD, FDX, FE, FISV, FITB, FRE,
 GENZ, GIS,
 HDI, HIG, HMA, HOT, HUM,
 JBL, JWN,
 INTU,
 KG, KMB, KMG,
 LH, LPX, LXK,
 MAT, MAS, MEL, MHS, MMM, MO, MVT, MX, MYG,
 NI, NKE, NTRS,
 PBG, PCAR, PFG, PGN, PNC, PX,
 RHI, ROK,
 SOV, SPG, STI, SUN,
 T, TE, TMO, TRB, TSG,
 UNP, UST,
 WHR, WY

\end{document}